\documentclass[twoside]{article}

\usepackage{lipsum} 
\usepackage{aas_macros}

\usepackage[sc]{mathpazo} 
\usepackage[T1]{fontenc} 
\usepackage{microtype} 

\usepackage[hmarginratio=1:1,top=32mm,columnsep=20pt]{geometry} 
\usepackage{multicol} 
\usepackage[hang, small,labelfont=bf,up,textfont=it,up]{caption} 
\usepackage{booktabs} 
\usepackage{float} 
\usepackage{hyperref} 
\usepackage{natbib} 
\usepackage{graphicx}
\usepackage{caption} 
\usepackage{subcaption} 
\usepackage{lettrine} 
\usepackage{paralist} 
\usepackage{wasysym}
\usepackage{stmaryrd}

\usepackage{abstract} 

\usepackage{color}

\usepackage{ulem}

\usepackage{titlesec} 
\renewcommand\thesection{\Roman{section}} 
\renewcommand\thesubsection{\thesection.\arabic{subsection}} 
\titleformat{\section}[block]{\large\scshape\centering}{\thesection.}{1em}{} 
\titleformat{\subsection}[block]{\large}{\thesubsection.}{1em}{} 

\usepackage{fancyhdr} 
\usepackage{color}

\title{\vspace{-15mm}\fontsize{24pt}{10pt}\selectfont\textbf{Vulcan: Retreading a Tired Hypothesis with the 2024 Total Solar Eclipse}} 

\author{
\large
\textsc{Michael B. Lund$^1$}\\
\normalsize $^1$Caltech/IPAC-NExScI\\
\normalsize \href{mailto:editor@actaprimaaprilia.com}{editor@actaprimaaprilia.com} 
\vspace{-5mm}
}
\date{}

\begin{document}

\maketitle 

\thispagestyle{fancy} 


\begin{abstract}

The number of planets in the solar system over the last three centuries has, perhaps surprisingly, been less of a fixed value than one would think it should be. In this paper, we look at the specific case of Vulcan, which was both a planet before Pluto was a planet and discarded from being a planet before Pluto was downgraded. We examine the historical context that led to its discovery in the 19th century, the decades of observations that were taken of it, and its eventual fall from glory. By applying a more modern understanding of astrophysics, we provide multiple mechanisms that may have changed the orbit of Vulcan sufficiently that it would have been outside the footprint of early 20th century searches for it. Finally, we discuss how the April 8, 2024 eclipse provides a renewed opportunity to rediscover this lost planet after more than a century of having been overlooked.
\end{abstract}


\begin{multicols}{2} 

\section{Introduction}
\lettrine[nindent=0em,lines=3]{O}ur understanding of the solar system has, in a sense, been in frequent flux for nearly a quarter of a millennium, starting with the discovery of the first planet not known since antiquity, Uranus in 1781\citep{Herschel1781}. Since then, our understanding of the solar system has grown with the discovery of new planets, such as Ceres \citep{Piazzi1801}, Vesta \citep{Olbers1807}, Neptune \citep{Galle1846}, Pluto \citep{Slipher1930, Shapley1930}, Sedna \citep{Brown2004}, and Eris \citep{Brown2005}. More controversial has been the cessation of some of those same planets that have been added to the solar system in the last 200 years \citep{Dymock2006, Lund2015}. Even now, additional candidates lurk on the outer edges of our solar system, most often under the Planet Nine or Planet X moniker \citep{Forbes1880, Lowell1915, Batygin2016}. Many other candidates have gone by other names, however, often much closer to home, and include candidates or former planets like Antichthon \citep{Aristotle350}, Neith \citep{Houzeau1884}, Phaethon \citep{Kugler1927}, Theia \citep{Daly1946}, Krypton \citep{Ovenden1972, Siegel1938}, and Tyche \citep{Matese1999}. Indeed, some efforts to identify planets even used up large swaths of the alphabet due to how many planets were being proposed, such as Pickering's planets O through U \citep{Hoyt1976} or Jones' planets A through Z \citep{Jones1953}.

Very few of these planets are linked to observations, and fewer still are linked to observations that can be relatively easily carried out. In this sense, planetary observations are in stark contrast with the scale, majesty, and ease of observation of solar eclipses, one of the most dramatic movements of the cosmic ballet going on \citep{Obrien1993}. While descriptions of additional planets are generally not recorded until the 18th century, records of solar eclipses at least go back to a Assyrian description of a solar eclipse from either 1375 BC \citep{Stephenson1970} or 1223 BC \citep{deJong1989}, over three millennia ago. There are Chinese documentations of lunar eclipses from roughly this same era \citep{Dubs1947}, as well as solar eclipses that are only slightly more recent \citep{Liu2003}, and some arguments to be made that oral records of eclipses can be matched back to nearly 5,000 years ago \citep{Masse1998}. From the time these eclipses were first documented to the present day, eclipses have remained for observation \citep{Simon1972} and provided unique opportunities \citep{Oz1986}.

Sitting at the intersection of the easy-to-observe solar eclipses and the harder-to-observe planets is the planet Vulcan, a case which we believe is worthy of additional consideration and deserving of a modern reassessment. Very few candidate planets have ever crossed one particular threshold, that of having been observed. More literally, few (if any) of them have crossed the more specific threshold of the limb of the sun. However, the existence of Vulcan was prompted initially by mass measurements but then was subsequently verified by transits. In Section~\ref{History} we outline the specific history of Vulcan from initial proposal in the 19th century to unceremonious disposal in the early 20th century (although it still mattered for the IAU\footnote{\url{https://www.iau.org/news/pressreleases/detail/iau1303/}} when it came time to name some of Pluto's moons as recently as 2013 \citep{Green2013}). In Section~\ref{Proposal} we look at what we can learn from astrophysical advances in the past century, particularly the rich diversity of exoplanets, and how those lessons might help us re-evaluate Vulcan. We discuss the role that the April 8, 2024 eclipse can play in this reassessment in Section~\ref{Eclipse} and then summarize this paper in Section~\ref{Summary}.

\section{Historical Context} \label{History}

\subsection{Vulcan Pre-history (before 1849)}

Vulcan's history begins not with its own discovery, but rather with that of Uranus in 1781 \citep{Herschel1781}. Alexis Bouvard had published position tables for Jupiter and Saturn in 1808, however at that point in time he did not include the relatively recently discovered Uranus \citep{Bouvard1808}. However, by 1822 he had been able to gather both observations that had been taken after the discovery of Uranus and earlier positions of Uranus that had been recorded by several astronomers that observed it without identifying it as a planet, namely Flamsteed, Mayer, Bradley, and Lemonnier \citep{Bouvard1822}. In trying to solve for an orbit, however, Bouvard discovered that there was no suitable way to reconcile the recent observations of Uranus with the older, pre-discovery observations that had been recovered. This led Bouvard to suggest that the issue was not with the accuracy of the observations but that something not yet detected was acting upon Uranus \citep{Bouvard1822b}.

While Bouvard was not able to fully complete this analysis before he died, this work was continued by others. Adding much more specificity to the question, independent predictions of a planet orbiting beyond Uranus were made by John Couch Adams and Urban Le Verrier based on the discrepancies in the motion of Uranus \citep{Adams1846, LeVerrier1846}. Following a request that Adams made for additional data made to British Astronomer Royal George Airy, there was also an attempt to locate a planet carried out by James Challis at Cambridge Observatory; while he did observe Neptune, he failed to identify it as such and indicated that the issue was insufficient rigor (noting that it would have been found if more stars had been compared but partially blaming the distraction of his cometary work) \citep{Airy1846, Challis1846}. This failure proved the opportunity for Neptune to instead be discovered by Johann Galle at the Berlin Observatory \citep{Galle1846} (the lack of coordination makes an early case for a more collaborative approach and the value of communication and coordination in follow-up observation programs \citep{Akeson2019}).

From here, interest was able to shift from the outermost regions of the solar system to the innermost regions of the solar system, and the difficulties that had been presented by Mercury, which began before the discovery of Uranus and continued after the discovery of Neptune. While its close proximity to the sun made observations of Mercury challenging, Johannes Kepler had still able to include Tycho Brahe's observations of Mercury in the Rudolphine Tables with sufficient precision that in 1629 they could be used to predict a transit of Mercury in 1631 \citep{Kepler1627}. While Kepler did not live to see how those predictions would play out (not entirely unlike Bouvard), the first transit of Mercury would be observed on November 7, 1631 by Pierre Gassendi; this observation yielded a radius of Mercury that was surprisingly small and prompted much discussion \citep{Gassendi1632, Hortensius1633, vanHalden1976, Police1983}. These transits would remain sufficiently important that the May 5, 1832 transit of Mercury would see a significant number of observations globally, such as Professor Uylenbroek and Mr. Kaiser in Leyden \citep{MNRAS1832d}, M. Quetelet at the observatory of Brussels, Mr. Henderson at the Cape of Good Hope, Captain Bayfield, R.N. in Quebec City, Captain R. Owen, R.N. in the Bahamas, and Captain Belcher, R.N. in modern-day Guinea-Bissau \citep{MNRAS1832a}, the Rev J. Fisher in Lisbon \citep{MNRAS1832b}, and Mr. Simms of the Royal Observatory at Greenwich and Mr. Riddle at Greenwich Hospital \citep{MNRAS1832c}.

The growing data on Mercury, and the sort of success that had come from using studies of Uranus' motion to find Neptune, would combine when Le Verrier turned his attention to the motion of Mercury, concluding that Mercury's peculiar motion could not be explained by accounting for known gravitational influences, but could be explained by a mass interior of Mercury's orbit and on a similar orbital plane \citep{LeVerrier1859}. The inner solar system had just become more crowded.

\subsection{Vulcan history (1849-1909)}

For a planet that would be this far inside Mercury's orbit and where the sun's glare would be such a challenge for detection, there were only two real detection methods, as noted by \citep{Campbell1909}:
\begin{quote}
Two special methods of discovery were applicable : (1) To
detect the planet projected upon the Sun's disk when its
orbital motion carried it between us and the Sun; (2) to
search for it when the sky background was darkened at the
time of a total solar eclipse.
\end{quote}
Of these two methods, the former was much more feasible to carry out, as solar eclipses are a very limited opportunity. Edmond Modeste Lescarbault, an amateur astronomer in France, had been inspired by his observations of the May 8, 1845 transit of Mercury that he had witnessed to search for intramercurial planets in a similar fashion. On March 26, 1859 he observed an object transiting in front of the sun, with a defined disk that was less than a quarter of the size he had observed for Mercury \citep{Lescarbault1860}. Le Verrier, in that same work, noted that based on the available observations of it, a circular orbit would place it at a semi-major axis of 0.1427 AU and so the planet would never be more than $8^{\circ}$ from the sun. It would be much fainter than Mercury (and only about one seventeenth the mass), such that the lack of detection thus far was hardly surprising. It should be noted, however, that even at this point the discovery was contested by observations made in Brazil by Emmanuel Liais \citep{Liais1860}. Lescarbault's observation would provide enough to stimulate a recalculation of the orbit of Vulcan by French astronomer Rodolphe Radau \citep{Radau1861}.

Additional observations of Vulcan would continue to be reported, however. Some, unfortunately, were reported with signifcant delays and were of limited value. F.A.R. Russell didn't report until 1876 his observation of what he believed to be an intramercurial planet transiting the sun that he observed on January 29, 1860 \citep{Russell1876}, though its publication in \textit{Nature} is a testament to its validity. A similarly delayed account appeared in \textit{Scientific American}, as it was observed by Richard Covington in around 1860, but he had presumed it to be a known object at the time and only reported it much later \citep{Covington1876}.

The next observations to be reported promptly would not come until 1862. On March 20th of that year, another amateur astronomer, W. Lummis of Manchester, observed the planet moving rapidly across the disk of the sun \citep{Hind1862}. Additional contemporary observations would follow, such as Aristade Coumbary's observations from Constantinople on May 8, 1865 \citep{Coumbary1865}.

The other approach, of observations during a solar eclipse, would not see much progress until the solar eclipse of July 29, 1878 that passed through the United States. During the eclipse, the discovery of a new planetary object was independently reported by two experienced professional astronomers. The first was by James C. Watson of Ann Arbor Observatory (who had previously discovered the asteroid 139 Juewa while in China to observe the 1874 transit of Venus \citep{Watson1874, Watson1875}), and whose eclipse expedition was mounted with the explicit goal of detecting intra-mercurial planets \citep{Watson1878a, Watson1878b}. The second was from Lewis Swift (perhaps best known for the comets he discovered), who observed the eclipse from Denver and also reported a sighting of Vulcan \citep{Swift1878a, Swift1878b}. While his report received some criticism \citep{Peters1879}, he would further explain and defend his report in subsequent publications \citep{Swift1878c, Swift1879}.

Photographic searches for intramercurial planets did not really come to maturity until the total solar eclipse of May 28, 1900. As part of the Harvard Observatory Expedition and their goal to use photography to search for intramercurial planets, Pickering discussed how their search involved an observing plan designed to minimize the confounding impact of sky brightness \citep{Pickering1900}, although when it came to the observations themselves, the telescope was jarred such that "all the plates taken with it were a total failure" \citep{Pickering1900b}. Similar adjustments were described by W.W. Campbell as a part of Lick Observatory's Crocker Expedition of 1900 \citep{Campbell1900}. In short, the photographic techniques had improved, but there hadn't yet been sufficient observations taken for anyone to feel like conclusions could be made regarding Vulcan.

On January 3, 1908, a total solar eclipse was visible in band through the South Pacific, and the Crocker Expedition from Lick Observatory was able to observe the moment of totality (during a very fortunate break in the rain that continued until only minutes before totality began) from Flint Island, 450 miles to the northwest of Tahiti at a latitude of $11^{\circ}$ south of the equator \citep{Campbell1908}. While over 500 stars were identified in the images taken, down to 9th magnitude, the point is noted that the field of view was $9^{\circ} \times 29^{\circ}$ oriented in line with the sun's equator, and so planets too inclined may have been missed by such a set of exposures \citep{Campbell1909}. As such, Campbell considered these observations "the closing of a famous astronomical problem." 

\subsection{Relativity Recently}
Solar eclipse observations would have a much more significant impact a decade after Campbell concluded that they had ruled out the presence of Vulcan in the most likely orbits. As Vulcan was falling out of favor, new physics was proposed that would impact both Vulcan and the sun. Einstein's 1905 paper on Special Relativity was groundbreaking \citep{Einstein1905}. In it and follow-up papers he addressed how gravity would impact light \citep{Einstein1908}. He later revisited this in \citet{Einstein1911}, looking at the amount that gravity would deflect light during a solar eclipse. This set the foundation for a planned attempt to look for this effect as early as the October 10, 1912 total solar eclipse but inclement weather prevented any conclusive results \citep{Perrine1923}. A later effort, for the May 29, 1919 solar eclipse, would prove much more successful and its measurements of light being bent by solar gravity would provide an experimental confirmation of relativity \citep{Dyson1920}. Confirmation would follow with Lick Observatory observations of the September 21, 1923 total solar eclipse \citep{Campbell1923}.

Somewhat in parallel, in 1915 Albert Einstein published his Theory of General Relativity \citep{Einstein1915a}. The same year, he also applied this to the observed precession of Mercury, largely explaining its movements but not explaining the observations that had been made of Vulcan \citep{Einstein1915b}. In effect, the radius of Vulcan was not impacted but the mass had been reduced, equivalent to a significant reduction of the density of Vulcan. Put another way, information had been gained that the initial assumptions of the composition of Vulcan were incorrect. In this sense, it is not at all unlike realizations that Pluto was not responsible for the variability in Neptune's orbit, but still existed \citep{Standish1993}.

\section{Proposal} \label{Proposal}

We are left, then, with one outstanding question. What happened to Vulcan between the late 1870s, when it was still being observed, and the 1900s when it was not detected photographically? While the destruction of Vulcan has been speculated on \citep{Abrams2009}, a planet can't simply disappear or be destroyed without notice \citep{Adams1979, Lund2017}. Collisions between planets are known to occur and have been observed \citep{Kenworthy2023, Marshall2023}, but a collision would likely be a noticeable event. It also is now documented that minor planets can go through a disintegration process close to a star and break up into smaller planetesimals as in WD 1145+017 \citep{Vanderburg2015}. However, this is a process that would have required Vulcan to be much closer in (Vulcan was on a period of 15-20 days compared to the objects in the WD 1145+017 system having periods of around 5 hours) and still appears to take place over longer timescales than the gap in Vulcan observations. This suggests that the outright destruction of Vulcan is unlikely.

\subsection{Vulcan Preservation}
However, another option remains that would explain not just why Vulcan was not observed by the cameras in 1908, but also why it has been absent from more recent observations of the sun's surface, as modern transits would surely be detected. Simply put, Vulcan is no longer on the orbit that it was as of the middle of the 19th century, but is on an inclined orbit that would have placed it close to the sun, but outside of the narrow band aligned with the sun's equator that \citet{Campbell1909} focused on. We suggest two possible causes for this that may have been overlooked by earlier researchers.

\subsubsection{Gravitational Scattering} \label{Scatter}
The first possibility, and the more straight forward one, is that Vulcan underwent a close gravitational interaction that significantly changed its orbit somewhere between roughly 1880 and 1900. Observers at the time, still working with a larger mass than Vulcan would later be found to have, would not have considered this without the presence of a massive object such that it would have been clearly detected before it had gotten too close to the sun. This mass threshold is much lower, however, once Vulcan is no longer constrained to something like 6\% the mass of Mercury as it was by \citet{Lescarbault1860}.

In looking at that window of time, there is one clear suspect body that both entered the inner solar system far enough to influence Vulcan, and also showed clear signs of gravitational strain. On September 7, 1882, Mr. Finlay, Chief Assistant at the Royal Observatory, Cape of Good Hope, was the first to spot what would become known as the Great Comet of 1882 as it was still approaching the sun \citep{Gill1882}. Observations of the comet as it continued to approach the sun were indicative of a single nucleus, up through its transit of the sun on September 17 \citep{Finlay1882, Elkin1882}. After transit, observers started to report that the nucleus of the comet was no longer consistent with a single object as it previously had been and appeared elongated by September 27 \citep{Barnard1882} or September 30 \citep{Gill1883}. The comet's nucleus had broken up \citep{Plummer1889}, and on October 17, 5 distinct fragments of the nucleus were identified \citep{Gill1883}.

We suggest, then, that our first possibility is that this fragmentation was a direct result of the Great Comet of 1882 passing close to Vulcan. We further suggest that this encounter also redirected Vulcan onto a highly inclined orbit. As the mass of Vulcan was believed to be higher at the time, this scenario would likely not have been considered.

\subsubsection{von Zeipel-Lidov-Kozai Mechanism}
The von Zeipel-Lidov-Kozai mechanism describes how two objects in orbit around one another can be influenced by a third, more distant object, particularly by exchanging inclination and eccentricity and takes place across a range of astronomical systems. The first work on this topic was published in 1910 by Swedish astronomer Edvard Hugo von Zeipel \citep{vonZeipel1910}, however his contributions did not receive proper attention until quite recently \citep{Ito2019}. Instead, this mechanism would not see much visibility in the broader community until two other scientists conducted independent (both with respect to each other and with respect to von Zeipel) work on this topic in the early 1960s, Michail L'vovich Lidov \citep{Lidov1962} and Yoshihide Kozai \citep{Kozai1962}. This delay in broader appreciation meant that the von Zeipel-Lidov-Kozai (Kozai hereafter) mechanism was not a concept the astronomical community broadly was familiar with until at least the 1960s, well after Vulcan was still an object of active investigation. For example, these sorts of mechanisms can result in planets in multiplanet systems developing high orbital inclinations \citep{Chatterjee2011} or even end up on retrograde orbits \citep{Naoz2011}, or create a highly eccentric close-in planet like HD 80606 \citep{Pont2009}. In other words, the highly-inclined orbit could be a result of the long-term impact of Kozai cycles caused by Mercury (discussed below) rather than by a single dramatic interaction as discussed in Case~\ref{Scatter}.

The timescale of these Kozai cycles is given in \citet{Merritt2013} as the following, where the subscript 2 corresponds to the outermost object:
\begin{equation}
\label{eq:TKozai}
T_{Kozai} = 2 \pi \frac{\sqrt{G M}}{G m_{2}} \frac{a^{3}_{2}}{a^{3/2}} (1 - e^{3}_{2} )^{3/2}
\end{equation}
This is presented with a simplified form as well:
\begin{equation}
\label{eq:TKozai_simplified}
T_{Kozai} = \frac{P^{2}_{2}}{P} (1 - e^{3}_{2} )^{3/2}
\end{equation}

In the case of Vulcan, we already have that its period was 19.7 days \citep{Radau1861}, and the closest perturber would be, slightly ironically given the original motivation that led to the discovery of Vulcan, the innermost planet Mercury. Mercury has an orbital period of 88.0 days and an orbital eccentricity of 0.206\footnote{\url{https://nssdc.gsfc.nasa.gov/planetary/factsheet/mercuryfact.html}}. Using these values in Equation~\ref{eq:TKozai_simplified}, we get that the timescale of the Kozai cycles for Vulcan would be 368 days, or just slightly longer than an Earth year. So this gives us a Kozai timescale that is considerably shorter than our 1880 to 1900 window during which Vulcan would need to undergo a significant orbital change.

This provides the alternative possibility that through the von Zeipel-Lidov-Kozai mechanism, Mercury was able to significantly excite the orbital inclination of Vulcan, resulting in Vulcan rarely being in line with the sun's equator and outside the region that intramercurial planet searches had deliberately targeted.

\subsection{April 8, 2024 Eclipse} \label{Eclipse}
We are now on the precipice of a unique opportunity to correct an oversight that has persisted for over a century. On April 8, 2024, a total solar eclipse will pass across North America, from Mazatlan in Mexico to Bonavista in Canada\footnote{\url{https://eclipsewise.com/2024/2024.html}}. Over a century ago, the search for intramercurial planets focused on a rectangular field in line with the sun's equator. Now, as we get yet another opportunity to witness a total solar eclipse, we suggest that while taking in the wonder that is a total eclipse, that observers also take a chance to look slightly above and slightly below the eclipse, outside the region that was previously searched for Vulcan. This is our next best opportunity to restore Vulcan to the pantheon of planets (or perhaps more realistically, dwarf planets) that make up our solar system.

\section{Summary}\label{Summary}
In this work, we have recounted the history of the intramercurial planet Vulcan, from when it was first proposed, to its history of observations, to its abrupt dismissal, as well as the astronomical and astrophysical context of these events. We have also described how either gravitational scattering or the von Zeipel-Lidov-Kozai mechanism may be responsible for causing Vulcan's orbit to change from its presupposed circular, coplanar orbit to a highly inclined orbit. This change in orbit would explain why Vulcan was not observed after the late 19th century. Finally, we highlight that the upcoming total solar eclipse provides an opportunity to rediscover Vulcan by looking for it close to the sun in regions that had previously been excluded from consideration when changes in orbital mechanics had not been factored in to the search.

\section{Acknowledgements}
The author wishes to thank Rob Siverd for feedback on this manuscript.

This research has benefited from the comprehensive biographies available at MacTutor, created and maintained by Edmund Robertson and John O'Connor of the School of Mathematics and Statistics at the University of St Andrews.


This research has made use of NASA’s Astrophysics Data System.


\bibliographystyle{apalike}
\bibliography{main}


\end{multicols}

\end{document}